\documentclass[aps,showkeys,superscriptaddress]{revtex4-2}

\usepackage{lipsum}
\usepackage{graphicx}
\usepackage{amsmath}
\usepackage{amssymb}
\usepackage{floatrow}
\usepackage{subfig}
\usepackage{xcolor}
\definecolor{mypurp}{RGB}{148,0,211}
\usepackage{soul}
\usepackage[singlelinecheck=off, justification=justified]{caption}

\begin{document}


\title{Collision of Dynamic Jamming Fronts in a Dense Suspension}


\author{Olav R\o mcke}
\affiliation{Norwegian University of Science and Technology, Department of Energy and Process Engineering, NO-7491 Trondheim, Norway}
\author{Ivo R. Peters}
\affiliation{University of Southampton, Faculty of Engineering and Physical Sciences, Highfield, Southampton SO17 1BJ, UK}
\author{R. Jason Hearst}%
\email{jason.hearst@ntnu.no}
\affiliation{Norwegian University of Science and Technology, Department of Energy and Process Engineering, NO-7491 Trondheim, Norway}



\date{\today}

\begin{abstract}
Dynamic jamming is a phenomenon whereby a dense suspension switches from a fluid-like to a solid-like state when subjected to sufficient stress and deformation. Large enough systems show that this transition is accompanied by a distinct jamming front. We present an experimental study where two jamming fronts are created simultaneously using two cylinders moving in parallel. We focus our observations on the collision of the jammed regions when the two fronts meet. Surprisingly, our measurements, combining surface texture visualization and time-resolved particle image velocimetry, show the formation of an unjammed region contained within the otherwise jammed suspension.
\end{abstract}



\maketitle

\section{\label{sec:Introduction}Introduction}


Densely packed suspensions of solid particles exhibit phenomena such as discontinuous shear thickening \cite{Wyart2014, Barnes1989, Seto2013}, dilating particle structure \cite{Brown2012} and the ability to jam \cite{Liu1998, Liu2010, Brown2014}, that is, switch to a solid-like state. In some systems, the transition from a fluid-like to a solid-like state is observed to give rise to the phenomena of dynamic shear jamming fronts \cite{Waitukaitis2012, Han2016, Han2019, Peters2016, Han2018, Peters2014, Romcke2021}. A region of high shear rate, initiated from a perturbing body, propagates through the suspension and leaves a jammed state in its wake. Although there has been recent progress in macro scale continuum modeling capturing dynamic shear jamming fronts \cite{Baumgarten2019, Han2019b}, studies of this phenomena have typically been limited to relatively simple flows and geometries. More specifically, previous experimental configurations generated jamming fronts from a single perturbing body, and thus the idea of jamming fronts interacting has not been previously investigated. What happens when jamming fronts collide is an open question.


A well known example of a suspension capable of dynamic jamming is a mixture of cornstarch and water \cite{Fall2012, Hermes2016, Peters2016}. Such a suspension flows when subjected to low stress, but shear thickens or jams at high stress. Once the stress is released, the suspension relaxes back to a fluid-like state \cite{Brown2014, Brown2012, Maharjan2017}. Physically, the underlying mechanism causing discontinuous shear thickening and dynamic jamming is understood as a transition from viscous to frictional interaction between particles, which changes the effective jamming volume fraction and subsequently the viscosity at a given volume fraction \cite{Wyart2014, Mari2014, Pan2015}. Repulsive forces between particles establish an onset stress \cite{James2018, Guy2015}, below which the particles are separated by a lubrication layer. However, when the suspension is subjected to a sufficiently high stress, capable of overcoming the repulsive force, the particles are brought into a friction dominated contact network \cite{Wyart2014, Singh2018}, which greatly increases the resistance to flow. It has been shown that particle surface chemistry \cite{James2018}, shape \cite{Brown2011,James2019} and poly dispersity \cite{Guy2020} affect this behaviour. The framework presented by Wyart \& Cates \cite{Wyart2014} identifies that a dynamically jammed state is accessible in the range of volume fractions where flow is possible for viscous, but not for frictional interactions  \cite{Guy2015, Peters2016, Romcke2021}.


A finite amount of strain as well as stress is required to build a frictional contact network capable of resisting flow \cite{Han2019b, Baumgarten2019}. As a consequence, large systems do not homogeneously turn into a jammed state, but transition locally once the suspension has strained sufficiently. This is observed as a front of high shear rate originating from the perturbing source that propagates through the suspension, leaving a jammed region in its wake \cite{Peters2014, Waitukaitis2012, Han2016, Han2019, Peters2016, Han2018, Romcke2021}. The stress is applied by the local acceleration of the suspension \cite{Han2019b}, while an intrinsic onset strain is observed to accompany the liquid-solid transition \cite{Majumdar2017, Romcke2021}, depending on volume fraction \cite{Han2016, Han2018, Han2019b}. For cornstarch suspensions, this transient phenomenon has been observed in domains ranging from $~30$~mm \cite{Han2019} to $300$~mm \cite{Peters2014}, where the transition into the jammed state can propagate unimpeded through the suspension before confinement effects from solid boundaries influence the behaviour \cite{Brown2012, Brown2014}. It is worth noting that the jamming front phenomena has also been observed at a geological (kilometer) scale \cite{Peters2015}. These examples are larger than typical milimeter sized rheometer configurations, as it is necessary to have a domain large enough to observe the front propagation. In summary, dynamic shear jamming fronts are observed in large systems when the suspension has a high enough volume fraction and is subjected to a sufficient amount of stress and strain to establish a force bearing contact network.


As mentioned above, the local stress is an important factor in determining if the suspension jams. Though \citet{Han2019b} have demonstrated how the stress is related to speed of the jamming front in the case of simple shear, we have no rigorous way of determining the local stresses in the general case. Instead, the dilating property of dense suspensions will be utilized here. Dilatancy is the ability of dense granular structures to increase in volume when sheared. In a suspension, the expansion of the solid phase causes a suction in the liquid phase \cite{Brown2012, Brown2014, Majumdar2017, Jerome2016, Fall2012}. This can be observed at the free surface as a change in surface texture from reflective to matte as the liquid phase is sucked into the voids in the expanding granular structure \cite{Maharjan2021, Brown2011, White2010, Smith2010}. A connection between free surface texture change, dilation and jamming in dense suspensions has been previously observed \cite{Smith2010, Brown2012, Majumdar2017, Jerome2016, Brown2011, Maharjan2021}, even showing solid-like behaviour such as cracking \cite{Smith2010, Roche2013, White2010, Allen2018}, followed by a relaxation back to a fluid-like state \cite{Smith2010, Roche2013}. This visual change has also been shown to correlate with the stress response~\cite{Brown2012, Maharjan2021}. Thus, observing the surface texture directly can be a means of characterizing different regions of the flow that do not appear obvious from velocity field data alone.


While earlier studies on dynamic shear jamming fronts have been limited to a single propagating front and its interaction with a solid boundary, this study presents the first observations of colliding jamming fronts. The jamming fronts are generated by two cylinders moving in parallel. We show that after collision, a diamond-shaped region between the cylinders is formed that relaxes back to its quiescent state, while still being surrounded by jammed material.


\section{\label{sec:Experiment}Experimental procedure}

\begin{figure}
    \sidesubfloat[]{\includegraphics[width=0.3\textwidth]{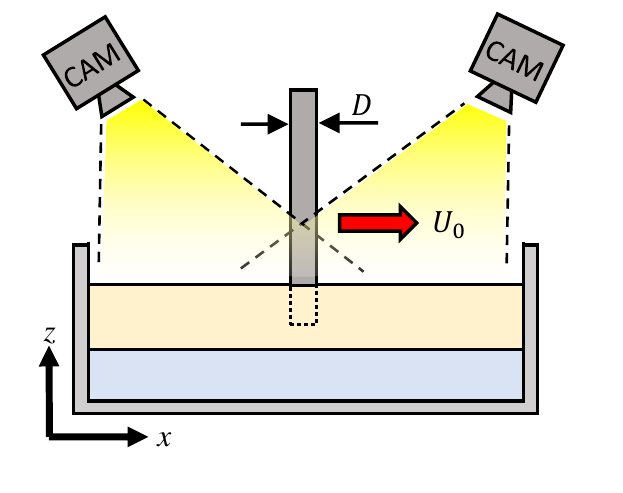}\label{figSetup:PIV}}
    \sidesubfloat[]{\includegraphics[width=0.3\textwidth]{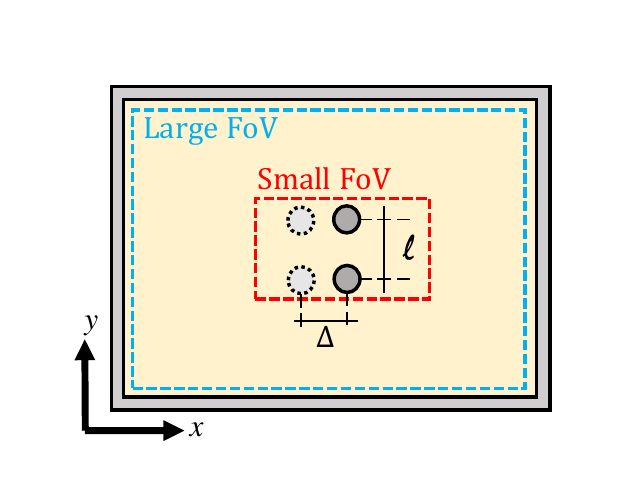}\label{figSetup:FoV}}
    \sidesubfloat[]{\includegraphics[width=0.3\textwidth]{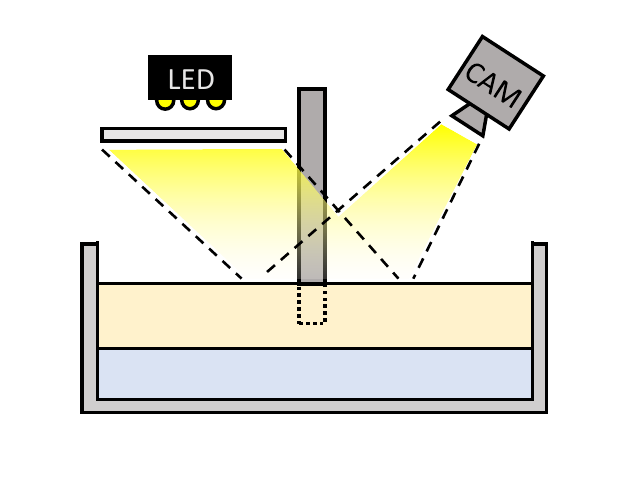}\label{figSetup:Tex}}
   \caption{Schematic of the experimental configuration. (a) PIV set-up with cylinder diameter ($D$) and direction and speed of travel ($U_0$) indicated. (b) The top view shows the two cylinders moving in parallel. The PIV fields of view are also indicated along with the cylinder position ($\Delta$) and separation ($\ell$). (c) Surface texture imaging configuration. The two side views (a) and (c) show the suspension floating on a layer of Fluorinert.}
   \label{figSetup} 
\end{figure}

Our experimental set-up, shown in Fig. \ref{figSetup}, is similar to that described in \cite{Romcke2021}. Experiments were conducted in a $1$~m$\times0.5$~m tank with two $25$~mm diameter ($D$) cylinders submerged into a $15$~mm thick suspension consisting of cornstarch and a sucrose-water solution. The suspension floated atop a $15$~mm layer of Fluorinert (FC-74). Fluorinert is a high density ($\rho=1.8$~g/ml), low viscosity ($\nu=0.75$~cSt) oil. This results in a close to stress free bottom boundary, such that the jamming front propagates in a 2D manner  \cite{Peters2014, Han2018, Romcke2021}.  Microscopically, the jamming mechanism is fully three-dimensional as the suspension layer thickness used in these experiments is several order of magnitude larger than the particle size. However, macroscopically, the jamming front propagation is approximately 2D in the suspension layer \cite{Peters2014, Han2018, Romcke2021}. Earlier lab scale observations of the jamming front phenomenon have been observed in domains up to $\sim300$~mm \cite{Peters2014}. The present configuration is more than three times larger than this, making it one of the physically largest lab scale experiments targeted at measuring shear jamming fronts, in order to reduce confinement effects.

\floatsetup[figure]{style=plain,subcapbesideposition=top}

The cylinder velocity ($U_0$) and the volume fraction ($\phi$) are chosen to ensure that we reach the dynamic shear jamming regime. The volume fraction is defined as
\begin{equation}
\label{eq:pf}
    \phi=\frac{(1-\beta)m_s/\rho_s}{(1-\beta)m_s/\rho_s + m_l/\rho_l + \beta m_s/\rho_w},
\end{equation}
where $m_s$ and $m_l$ are the starch and suspending fluid mass, respectively. $\rho_s=1.63$~g/ml and $\rho_w=1.00$~g/ml are the starch and water densities, respectively. The suspending liquid is a $50$\% wt. sucrose solution with density $\rho_l=1.23$~g/ml and viscosity $\eta_0=0.016$~Pa~s. The sucrose is added to the water in order to reduce the impact of settling \cite{Romcke2021}. Due to the difficulty in precisely determining the nominal value of the volume fraction of cornstarch suspensions, earlier studies have used measured quantities such as onset strain rate \cite{Maharjan2017} or front propagation \cite{Peters2014} as the control parameter rather than the volume fraction. We use the same approach here, as the behaviour was highly consistent within a batch of suspension. The discrepancy in the volume fractions was accounted for by a $2$\% variation in the starch water content, $\beta$, and was nominally $9\% \le \beta \le 11\%$. Thus, the $\phi$ reported here reflect the trend in the front propagation. At sufficiently low $U_0$ no propagating jamming front is observed \cite{Han2016, Romcke2021}. At sufficiently high perturbing speeds or stresses, yielding and buckling of the suspension have been reported \cite{Peters2014, Han2019b}. The results presented here are for $U_0=0.14$~m/s where the suspension is known to jam and the front propagation factor is independent of the $U_0$ \cite{Romcke2021}. Ancillary tests at $U_0 = 0.11$ and $0.18$~m/s confirm this.

The suspension is mixed for a minimum of two hours before it is poured onto the Fluorinert in the tank. It is then pre-sheared before any measurements are performed. The surface is seeded with black pepper and for every batch, the measurements are repeated 12 times. The time from when the suspension is poured until the experiment is complete and the suspension is discarded is approximately 15 min, which results in an estimated settling distance on the order of the particle diameter, and is thus negligible \cite{Garside1997, Richardson1997}. Two high speed cameras (Photron FASTCAM Mini WX100) view the suspension surface in front and behind the traversing cylinders (figure \ref{figSetup:PIV}). Particle image velocimetry (PIV) is used to converted the image series to velocity fields with LaVision DaVis 8.4.0. An initial pass was performed with $96$~pixels~$\times$~$96$~pixels square interrogation windows, followed by two passes with circular interrogation windows with decreasing size ending at $48$~pixels~$\times$~$48$~pixels. For all passes, the interrogation windows have a $50$\% overlap. The final velocity fields are made-up of the two fields, one from each camera, stitched together, to provide a complete view around the cylinders. PIV measurements were conducted at two magnifications using different lenses. A 50~mm lens was used to image the entire flow field around the pair of cylinders, and in a separate experiment a 180~mm lens was used to specifically focus on the area between the two cylinders. The resulting PIV fields of view are represented schematically in figure \ref{figSetup:FoV}.


Only minor adjustments to the set-up were needed to enhance the visibility of the surface texture. The surface texture measurement campaign was performed in order to provide a second method for identifying jammed regions in the flow. Here, an LED array covered by a semi transparent acrylic sheet served as the light source. The camera was positioned such that it captured the direct reflection of the light source in the suspension surface \cite{Maharjan2021}. A rough surface, caused by dilation, diffuses the light reflected off the suspension surface and thus changes the observed intensity. As such a matte surface indicates that the suspension is jammed while a reflective surface indicates that the suspension is in a quiecent unjammed state \cite{Brown2012, Brown2014, Jerome2016, Maharjan2021}. The size of the acrylic sheet ensured that the reflection in the suspension surface covered the field of view of the camera (Figure \ref{figSetup:Tex}). In order to get as clear a view of the surface as possible, we did not seed the sample with tracer particles for this experiment.


\section{\label{sec:Result} Results}


Both the cylinders and the suspension start at rest. The cylinders are initially located at $x=0$. Once cylinder movement is initiated, large velocity gradients are observed at the surface of each cylinder, which propagate into the surrounding suspension. At some point, depending on the speed of the front and the distance between the cylinders, $\ell$, the fronts collide. By the end of an experimental run, the region between the two cylinders moves at approximately the perturbing speed. In addition to the velocity field ($\mathbf{u}$), we also present the strain rate magnitude ($\dot{\gamma}$) and the accumulated strain ($\epsilon$). The strain rate magnitude is calculated as
\begin{equation}
    \dot{\gamma}=\sqrt{2\mathcal{D}_{ij}\mathcal{D}_{ij}}\text{, \ where \ }\mathcal{D}_{ij}=\frac{1}{2}\left(\frac{\partial u_i}{\partial x_j} + \frac{\partial u_j}{\partial x_i}\right).
    \label{eq:StrainRate}
\end{equation}
The accumulated strain ($\epsilon$) is calculated as the norm of the Eulerian logarithmic strain tensor \cite{Nasser2004, Romcke2021}. This is estimated from the velocity field by first calculating the movement of the material points as $\mathbf{x_p}(\mathbf{X},t)=\mathbf{X}+\int_0^t\mathbf{u}(\mathbf{x_p}(\tau),\tau)d\tau$, from which the deformation gradient tensor is acquired as $\mathbf{F}=\partial\mathbf{x_p}/\partial\mathbf{X}$. Here, $\mathbf{X}$ is the position of the material points at $t=0$. A polar decomposition of the deformation gradient ($\mathbf{F}=\mathbf{VR}$) gives the left stretch tensor $\mathbf{V}$, which has eigenvalues and eigenvectors, $\lambda_i$ and $\mathbf{n}_i$, respectively. The strain tensor has the spectral representation \cite{Nasser2004}
\begin{equation}
    \mathbf{e}=\sum_i \ln(\lambda_i) \mathbf{n}_i\otimes\mathbf{n}_i,
    \label{eq:Strain}
\end{equation} 
where we order eigenvalues $\lambda_1>\lambda_2$, such that the direction of stretch and compression is denoted $\mathbf{n}_1$ and $\mathbf{n}_2$, respectively. $\mathbf{n}_1$ and $\mathbf{n}_2$ are orthogonal. The norm of the strain tensor $\epsilon=||\mathbf{e}||$ is used as a scalar measure of the accumulated strain.

\begin{figure}
    \includegraphics[width=0.89\textwidth]{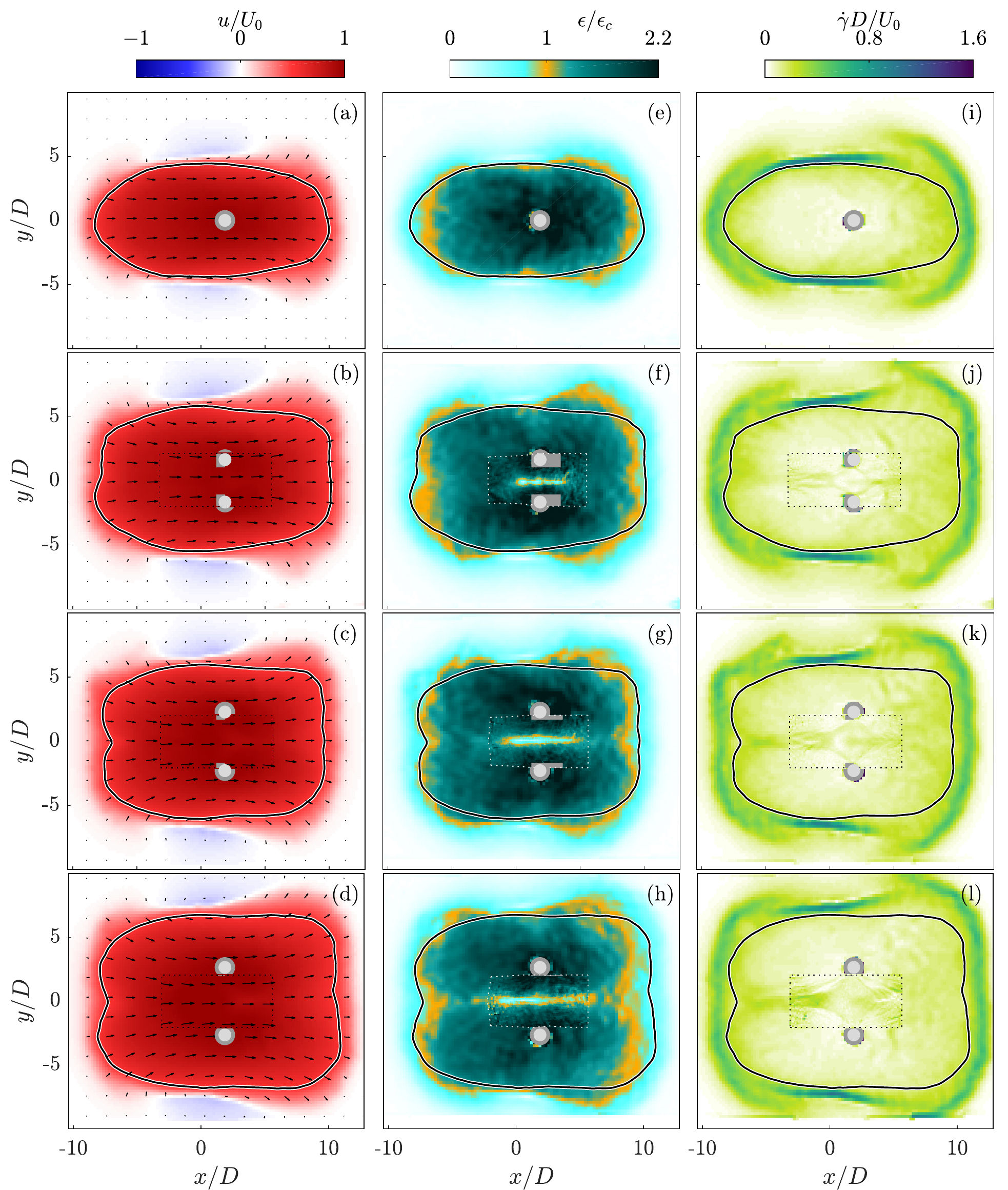}
   \caption{Snapshots of the flow field at $\Delta/D=1.87$ and $\phi=0.36$ for several cylinder separations. From top to bottom $\ell/D=0$, $3.3$, $4.5$ and $5.3$. Results from the small field of view are placed as an overlay in the center region indicated by the dotted border. The cylinders are moving from left to right. (a)-(d)~Velocity ($u$) with velocity vectors, indicating that the suspension is moving with the cylinder(s). (e)-(h)~Accumulated strain ($\epsilon$). The accumulated strain is normalized by the onset strain ($\epsilon_c$) accompanying the jamming front found to be $0.14\pm0.03$ from the single cylinder reference case at this specific volume fraction. (i)-(l)~Strain rate magnitude ($\dot{\gamma}$).  The light gray circles identify the cylinders. Dark gray regions cover the region obstructed from view by the cylinders themselves. For the accumulated strain, the impacted region is larger because we are unable to track the full deformation history due to the obstructions to the field of view.}
   \label{figA} 
\end{figure}

\begin{figure}
    \includegraphics[width=0.87\textwidth]{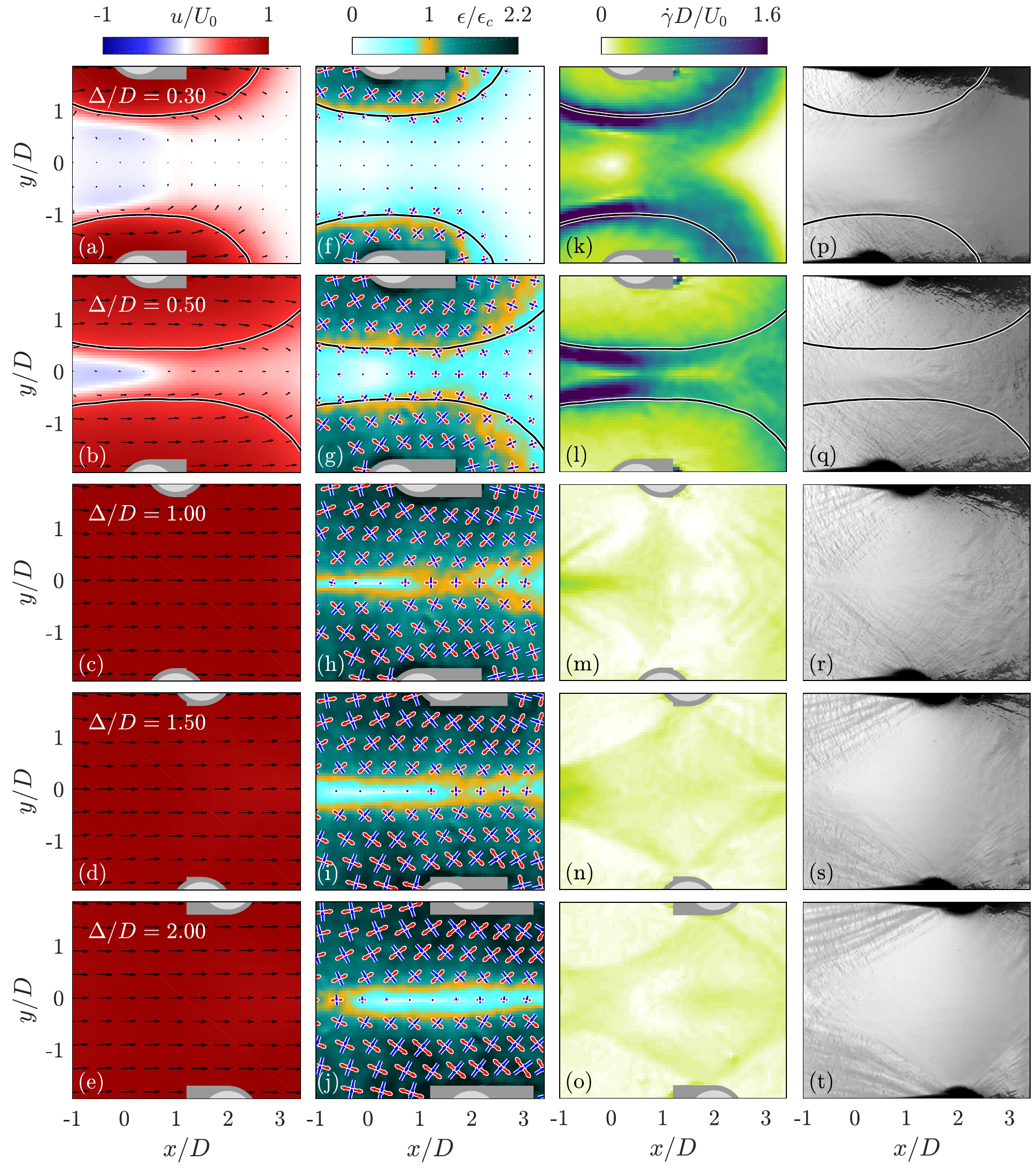} 
   \caption{Snapshots of the colliding jamming fronts for $\ell/D=4.5$ and $\phi=0.36$. The rows show different displacement of the cylinders, $\Delta/D$. (a)-(e) are the velocity fields. (f)-(j) are the accumulated strain, $\epsilon$, with superimposed eigenvectors, $\mathbf{n}_1$ (red) and $\mathbf{n}_2$ (blue), from eqn. \eqref{eq:Strain}. (k)-(o) are the strain rate, $\dot{\gamma}$, from eqn. \eqref{eq:StrainRate}. (p)-(t) show the result of the surface texture experiment. The black line superimposed on all fields represents the $0.5U_0$ contour from the PIV data, which is a surrogate for the jamming front. The contour lies outside of the presented fields for $\Delta/D\ge 1.00$. The masked regions are the same as in figure \ref{figA}.}
   \label{figB} 
\end{figure}

PIV results are presented in Fig. \ref{figA} which shows both the large and small fields of view at $\Delta/D=1.87$, which is close to the end of an experimental run. In particular we look at different cylinder separations ($\ell$) with the single cylinder case as a reference. The velocity vectors are plotted in figure \ref{figA}a-\ref{figA}d, and the $0.5U_0$ contour, typically used to identify the jamming front \cite{Waitukaitis2012, Peters2014, Peters2016, Han2016, Han2018, Han2019, Han2019b}, is indicated by the black line in all fields. Some aspects of the system behaviour are independent of both the number of cylinders and their separation. Relative to the perturbing body, the front has propagated twice as fast in the longitudinal direction compared to the transverse direction. This 1:2 propagation relationship is a direct consequence of an underlying onset strain \cite{Han2016, Romcke2021}, and does not change by introducing a second cylinder. A notable difference with the reference case, however, occurs in the collision region between the two cylinders. A uniform velocity field has in previous studies been used to indicate the jammed region \cite{Peters2014, Han2016}. Here, the velocity fields in Figs. \ref{figA}b-\ref{figA}d (seen in more detail in figure \ref{figB}c-\ref{figB}e)  might suggest a uniformly jammed suspension after the fronts have collided. Though the velocity is relatively uniform, the strain rate represented as $\dot{\gamma}$ in Figs. \ref{figA}j-\ref{figA}l, show traces of a diamond-like structure in the collision region. The accumulated strain ($\epsilon$) reveals a slightly different picture. As seen in figures \ref{figA}f-\ref{figA}h, $\epsilon$ indicates the presence of a slit of material that has not sufficiently deformed to transition into the jammed state. This suggests that there is a region between the two cylinders that is unjammed while the surrounding suspension is jammed. 

Fig. \ref{figB} compares PIV data with the high speed imaging of the suspension surface texture. The texture measurements have no direct information about the velocity field. However, as the example shown here is at the same volume fraction, cylinder configuration and speed, the $0.5U_0$ contour from the PIV data is superimposed on both PIV and texture images. At later stages in an experimental run (here, $\Delta/D\geq1.00$), the contour lies outside of the field of view shown in figure \ref{figB}. In the $\dot{\gamma}$ fields it can be seen that the $0.5U_0$ contour corresponds to a strong shear front propagating away from the cylinders as the flow evolves. After the two fronts collide, the suspension moves with the cylinders as indicated by the uniform velocity field (Fig \ref{figB}c-\ref{figB}e). The accumulated stain (figure \ref{figB}f-\ref{figB}j) shows the development of the undeformed slit. Interestingly, the region between the two cylinders is populated by a diamond-like shape visible in both the shear rate magnitude (figure \ref{figB}m-\ref{figB}o) and the surface texture images (figure \ref{figB}r-\ref{figB}t). Based on the surface texture, we infer that the suspension between the cylinders, which appears to be reflective, has relaxed back to a quiescent, unstressed state. Thus confirming that indeed the region between the cylinders is unjammed.


From previous studies where the front propagates freely through the suspension, it has been shown that the accumulated strain can be used to quantify when the suspension transitions into a solid-like state. However, in the collision region, the strain does not account for the subsequent relaxation of the suspension revealed by the texture images. As a consequence, there is a discrepancy between the shape of the unjammed region revealed by the accumulated strain and the texture images. Traces of the diamond shape observed in the texture image exist in the strain rate. This is presented in figure \ref{figB}. Though the accumulated strain ($\epsilon$) does not show a diamond shape as observed in the texture images, the eigenvectors of the strain tensor reveal an interesting feature. These eigenvector seem to roughly align with the boundary of the diamond shape (see figure \ref{figB}f-\ref{figB}j).

\begin{figure}
    \includegraphics[width=0.5\textwidth]{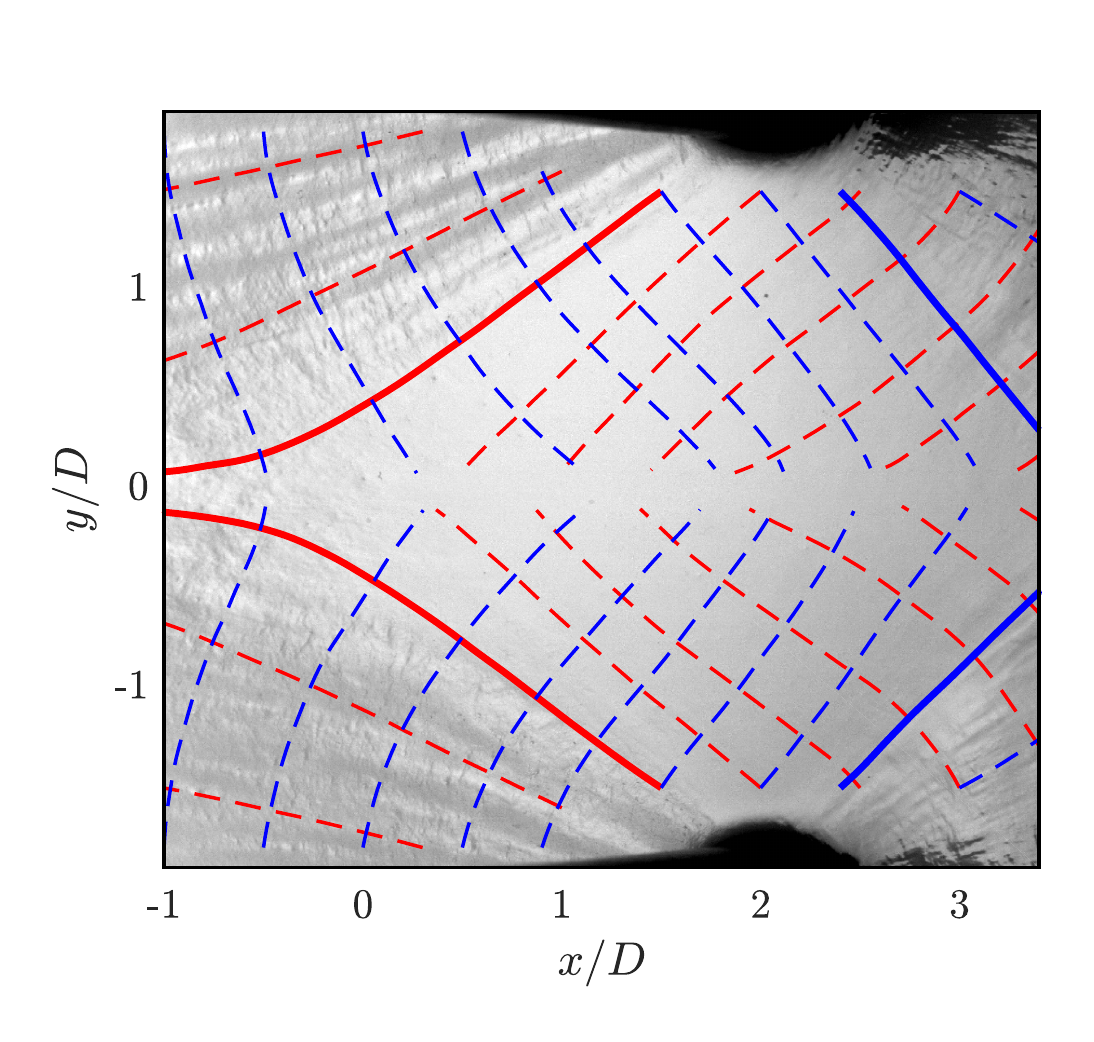}
   \caption{Lines of principle strain superimposed on the texture image at $\Delta/D=2.00$. Stretch ($\mathbf{x}_1$) and compression ($\mathbf{x}_2$) lines are represented by the red and blue lines, respectively. Here we have emphasized the lines that roughly outline the unjammed region in the texture image.}
   \label{figE} 
\end{figure}

We further investigate the apparent discrepancy by plotting the lines of principle strain \cite{Romcke2021} superimposed on a texture image in figure \ref{figE}. For an instantaneous strain field the principle strain lines are acquired numerically by solving $d\mathbf{x}_1/d\sigma = \mathbf{n}_1$ for principle stretch and $d\mathbf{x}_2/d\sigma = \mathbf{n}_2$ for principle compression, which ensures that the lines $\mathbf{x}_1(\sigma)$ and $\mathbf{x}_2(\sigma)$ are tangential to the eigenvectors $\mathbf{n}_1$ and $\mathbf{n}_2$, respectively. $\sigma$ is an arbitrary parametrization of the lines. Solving for $\mathbf{x}_1$ and $\mathbf{x}_2$ for multiple initial positions is presented in figure \ref{figE}. We avoid the center slit when calculating $\mathbf{x}_1$ and $\mathbf{x}_2$, as the accumulated strain is small in this region. Here we have highlighted the lines that roughly correspond to the diamond shape revealed by the texture image. Note that in the fore half plane, it is the compression line, $\mathbf{x}_2$, that outlines the diamond boundary, while in the aft half it is the stretch line, $\mathbf{x}_1$. 

As discussed in the introduction, both the stress and the accumulated strain need to be sufficiently high to achieve (or remain in) a jammed state \cite{Baumgarten2019, Han2019b}. The slit in the center revealed by the accumulated strain never transitions into a jammed state as the strain level is not sufficiently high. The subsequent relaxation of the surrounding diamond-shaped follows, as stresses are no longer transmitted through this region, revealed by the texture images.



\begin{figure}
    \includegraphics[width=0.5\textwidth]{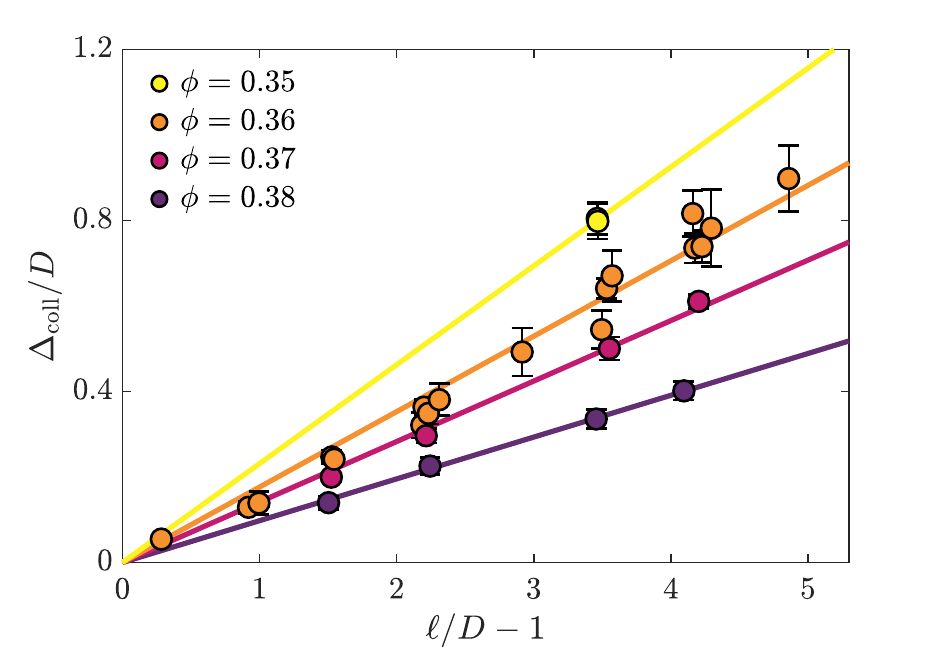}
   \caption{cylinder displacement at the time of front collision, $\Delta_{\mathrm{coll}}/D$, compared to the effective surface-to-surface cylinder separation $\ell/D-1$. Linear fits are meant only for reference. } 
   \label{figD} 
\end{figure}

Additional PIV measurements at different volume fractions within the dynamic jamming range confirm the same qualitative behaviour. A slit of material between the two cylinders does not deform the amount expected for jamming to occur even though the velocity field appears uniform. With increasing volume fraction, the accumulated strain needed for the jamming transition to occur decreases, while the front propagates faster through the suspension \cite{Peters2014, Peters2016, Han2018}. This is summarized in Fig. \ref{figD} showing how the moment of collision of the jamming fronts varies with volume fraction ($\phi$) and cylinder separation ($\ell$). We define the collision as the event when the jamming front transitions from separate contours around each cylinder to one contour enveloping both cylinders. This is quantified by the distance the cylinder has traveled from its starting point to when collision occurs ($\Delta_{\mathrm{coll}}$). The lines are a guide for the eye, but a linear trend is evident. The points are expected to fall onto a straight line as long as the relation between front and cylinder movement is constant, which Fig. \ref{figD} confirms.

\section{\label{sec:Conclusion} Conclusions}

We have presented the first observations of colliding jamming fronts. These fronts were generated by two cylinders moving in unison, and PIV has been used to calculate the velocity and deformation fields. The suspension between the cylinders moves uniformly after front collision. High speed observations of the free surface texture show that after collision a diamond-like region is formed between the cylinders which relaxes back to its quiescent, unjammed state. The accumulated strain does not portray this diamond shaped region. However, we show that the edges of the diamond shape revealed by the texture images align with the direction of principle strain. We attribute this apparent discrepancy between the two approaches to the fact that the accumulated strain does not account for the relaxation of the suspension.

Although the region between the cylinders appears unjammed, there still must be a force-bearing structure connected to the two cylinders to drive the jamming front that continues to propagate through the suspension away from the bodies.  This structure would effectively be protecting the region between the cylinders, allowing it to remain unjammed, and has a diamond-shape in the present configuration. By utilizing a combination of PIV and surface texture measurements we have thus identified the phenomena associated with colliding jamming fronts. Surprisingly, our results reveal that colliding jamming fronts unjam a region between the perturbing bodies, providing a new way geometry can be used to control where jamming occurs and where it does not.

\begin{acknowledgments}
RJH acknowledges the financial support of the Research Council of Norway (Grant No.~288046). IRP acknowledges financial support from the Royal Society (Grant No.~RG160089). Data supporting this study are openly available from the University of Southampton repository (DOI to be provided).
\end{acknowledgments}




\providecommand{\noopsort}[1]{}\providecommand{\singleletter}[1]{#1}%

\end{document}